\documentclass[12pt,aps,showpacs]{revtex4}
\usepackage{graphicx}

\begin{document}

\title{Quantum Dynamics of the Oscillating Cantilever-Driven Adiabatic
 Reversals in Magnetic Resonance Force Microscopy}
\author{G.P. Berman$^1$, F. Borgonovi$^2$, and V.I.
 Tsifrinovich$^3$}
\affiliation{$^1$ Theoretical Division, Los Alamos National
Laboratory,
Los Alamos, New Mexico 87545}
\affiliation{$^2$Dipartimento di Matematica e Fisica, Universit\`a Cattolica,
 via Musei 41 , 25121 Brescia, Italy, and I.N.F.M., Unit\`a
 di Brescia, Italy, and I.N.F.N., sezione di Pavia, Italy}
\affiliation{$^3$ IDS Department, Polytechnic University, Brooklyn, NY 11201}
\vspace{5mm}
\begin{abstract}
We simulated the quantum dynamics for magnetic resonance force
microscopy (MRFM) in the oscillating cantilever-driven adiabatic
reversals (OSCAR) technique.
 We estimated the frequency shift of the cantilever vibrations and
demonstrated that this shift causes the formation of a
Schr\"odinger cat state which has some similarities and
differences from the conventional MRFM technique which uses
cyclic adiabatic reversals of spins. The
 interaction of the cantilever with the environment is shown to quickly destroy the coherence
between the two possible cantilever trajectories. We have shown that using
partial adiabatic reversals, one can produce a  significant increase in the
OSCAR signal.
\end{abstract}

\maketitle
\section{Introduction}
The oscillating cantilever-driven adiabatic reversals (OSCAR)
technique is, probably, the most promising way to achieve
single-spin detection using magnetic resonance force microscopy
(MRFM) \cite{1,2}. In the OSCAR technique,
 the oscillating cantilever produces an oscillating
$z$-component of the effective magnetic field in the system of
coordinates connected to the rotating {\it rf} field, which causes
the magnetic moment reversals. In turn, the cyclic adiabatic
reversals of the magnetic moment cause the frequency shift in the
cantilever vibrations, which is to be measured.

While the classical dynamics of the OSCAR has been studied in
\cite{1,2}, the quantum dynamics has never been investigated. This
paper describes the quantum dynamics of the spin-cantilever system
in the OSCAR technique. In Section 2, we introduce the quantum
Hamiltonian and the equations of motion for the system. In Section
3, we present a qualitative analysis and estimates  of the OSCAR
signal. We show that the OSCAR signal can be increased by
sacrificing the full reversals of the effective field. Namely,
partial adiabatic reversals can provide a larger signal than the
full reversals. We estimate the maximum OSCAR signal generated by
a single spin. In Section 4, we present numerical simulations of
quantum dynamics using the Schr\"odinger equation. In particular,
we demonstrate the formation of a Schr\"odinger cat state which
simultaneously includes two possible trajectories of the
cantilever. In Section 5, we present numerical simulations of the
decoherence caused by the interaction of the cantilever with the
thermal environment.

\begin{figure}
\begin{center}
\includegraphics[width=11cm,height=7cm]{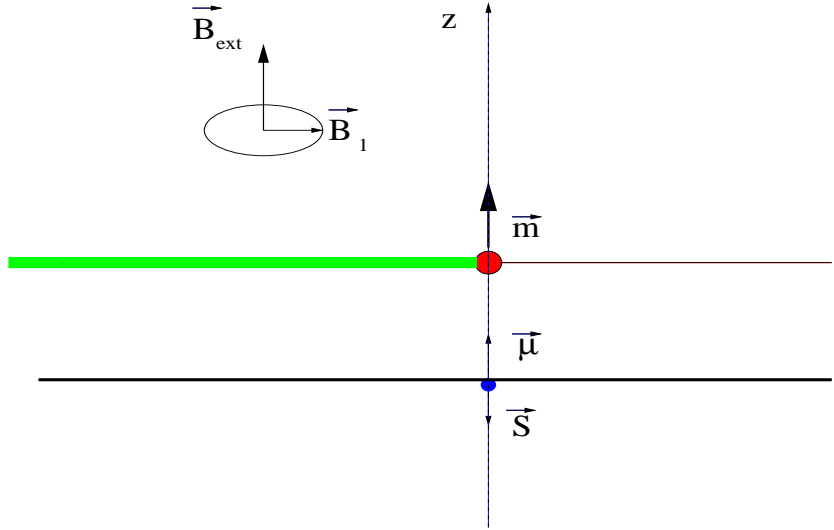}
\vspace{-5mm}
\caption{The geometry
of the OSCAR MRFM. ${\vec m}$ is the magnetic moment of the
ferromagnetic particle on the cantilever tip; ${\vec \mu}$ and
${\vec S}$ are the magnetic moment and the spin of an atom (in the
ground state); ${\vec B}_{ext}$ and ${\vec B}_1$ are the permanent
and rotating magnetic fields.} \label{uno}
\end{center}
\end{figure}

\section{The Hamiltonian and equations of motion}

We describe the cantilever interacting with a single electron spin
1/2 as a harmonic oscillator with the fundamental frequency
$\omega_c$. The dimensionless quantum Hamiltonian of this system
can be written (in the rotating coordinate systems) as
$$
{\cal H}=\frac{1}{2}(p^2+z^2)+\varepsilon S_x-2\eta zS_z, \eqno(1)
$$
with the dimensionless parameters
$$
p=P/P_0,~z=Z/Z_0,~\varepsilon=\gamma B_1/\omega_c,~
P_0=\hbar/Z_0,\eqno(2)
$$
$$
\eta=\gamma\hbar\Bigg|{{\partial B_z}\over{\partial
z}}\Bigg|/2(\hbar
\omega_ck_c)^{1/2},~Z_0=(\hbar\omega_c/k_c)^{1/2}.
$$
Here $P$ and $Z$ are operators of the momentum and the
$z$-coordinate of the cantilever, $k_c$ is the cantilever spring
constant, $B_1$ is the magnitude of
 the rotating magnetic field, $\gamma$ is the magnitude of the electron gyromagnetic ratio,
$B_z$ is the $z$-component of the magnetic field acting on the
spin, including the external permanent magnetic field $B_{ext}$
and the magnetic field produced by the ferromagnetic particle.
(See Fig. 1.)

The quantum dynamics of the MRFM was studied earlier in \cite
{3,4}.  In these papers the oscillating $z$-component of the
effective magnetic field acting on the spin was considered to be
generated by frequency modulation of the external rotating field
$B_1$. The $z$-component of the effective field, $-2\eta z$,
associated with the cantilever vibrations, was negligible in
comparison with the $z$-component of
 the external effective field. Below we consider the opposite (OSCAR)
 situation: the oscillating $z$-component of the effective field is
produced by cantilever vibrations.

To simulate the Schr\"odinger dynamics, we use the dimensionless
Schr\"odinger
 equation
$$
i{{\partial \Psi}\over{\partial\tau}}={\cal H}\Psi, \eqno(3)
$$
where the wave function $\Psi$ is
$$
\Psi=u_\alpha(z,\tau)\alpha+u_\beta(z,\tau)\beta, \eqno(4)
$$
$$
\alpha=\left(\matrix{1\cr 0\cr}\right),~\beta=\left(\matrix{0\cr
1\cr}\right),~\tau=\omega_ct.
$$
The orbital wave functions $u_{\alpha,\beta}(z,\tau)$ have been
expanded in the
 eigenfunctions, $u_n(z)$, of the oscillator Hamiltonian with time-dependent
coefficients. The resulting system of differential equations
for time-dependent coefficients was solved for a finite set of
basis functions $u_n(z)$. Alternatively, we have found eigenvalues
and eigenvectors of the time-independent Hamiltonian ${\cal H}$
for the same finite basis, and constructed from them the time
dependent solution. These two approaches have been shown to give
the same results.

To simulate the decoherence process, we used the simplest master
equation (an ohmic model in the high-temperature approximation
\cite{5})
$$
{{\partial\rho_{s,s^\prime}}\over{\partial\tau}}
=\Bigg[{{i}\over{2}}\Bigg( {{\partial^2}\over{\partial
z^2}}-{{\partial^2}\over{\partial z^{\prime 2}}} \Bigg)-
{{i}\over{2}}(z^2-z^{\prime 2})\eqno(5)
$$
$$
-{{1}\over{2Q}}(z-z^\prime) \Bigg({{\partial}\over{\partial
z}}-{{\partial}\over{\partial z^\prime}}
\Bigg)-{{D}\over{Q}}(z-z^\prime)^2
$$
$$
 -2i\eta(z^\prime s^\prime-zs)\Bigg]\rho_{s,s^\prime}-
i{{\varepsilon}\over{2}}(\rho_{s,-s^\prime}-\rho_{-s,s^\prime}),
$$
where $\rho_{s,s^\prime}\equiv\rho_{s,s^\prime}(z,z^\prime,\tau)$
is the density matrix, $s,s^\prime=\pm 1/2$,
$D=k_BT/\hbar\omega_c$ is the dimensionless diffusion coefficient,
and $Q$ is the quality factor of the cantilever.

The function $\rho_{s,s^\prime}(z,z^\prime,\tau)$ was expanded in
the basis of the eigenfunctions $u_n(z)u^*_m(z^\prime)$ with
time-dependent coefficients.
 These coefficients were calculated numerically for a finite set of basis
functions.

\section{Qualitative analysis and estimates of the OSCAR signal}

Suppose that initially a single spin is parallel to the effective
magnetic field ${\vec B_{eff}}=(\varepsilon,0,-2\eta z)$. Under
the conditions of
 adiabatic motion, the spin component along the effective magnetic field
is an integral of motion. Thus, in the process of motion the spin
will have the same definite direction relative to the effective
magnetic field. Consequently, the orbital and spin degrees of
freedom are not entangled: the wave function of the system will
remain a tensor product of the orbital and spin parts. As a
result, the spin-cantilever correlators are equal to zero, and the
Heisenberg equations of motion
$$
\dot z=p,~\dot p=-z+2\eta S_z,\eqno(6)
$$
$$
\dot S_x=2\eta zS_z,~\dot S_y=-2\eta zS_x-\varepsilon S_z,~\dot
S_z=\varepsilon S_y,
$$
are equivalent to the classical equations of motion. That is why
in this case we can use classical estimates for the single-spin
OSCAR signal.

We estimate the OSCAR signal in the following way: under the
conditions of adiabatic motion (and if the spin is parallel to the
effective field) we have
$$
S_z/S=\mp B_{eff}^z/B_{eff},\eqno(7)
$$
where the upper sign corresponds to the ground state (the spin
points in the direction opposite to the direction of the effective
field). It follows from Eq. (7) that
$$
S_z=\pm{{\eta z}\over{\sqrt{(2\eta z)^2+\varepsilon^2}}}.\eqno(8)
$$
Now, we substitute this expression into the equation for the
cantilever coordinate $z$
$$
\ddot z+z-2\eta S_z=0.\eqno(9)
$$
Neglecting oscillations with twice the  cantilever frequency, we
have from Eqs. (8) and (9)
$$
\ddot
z+z\mp{{2\eta^2z}\over{\sqrt{2\eta^2z_m^2+\varepsilon^2}}}=0,\eqno(10)
$$
where $z_m$ is the amplitude of the cantilever vibrations.
Assuming that the cantilever frequency shift $\Delta\omega_c$ is
small in comparison to the unperturbed cantilever frequency,
$\omega_c$, $(|\Delta\omega_c|\ll\omega_c)$ we have the estimate
for the OSCAR signal:
$$
{{\Delta\omega_c}\over{\omega_c}}=\mp{{\eta^2}\over{\sqrt{2\eta^2z^2_m+
\varepsilon^2}}}.\eqno(11)
$$

Note that the frequency shift (11) decreases with increasing of
the cantilever amplitude $z_m$. The reason for this dependence is
the following. The frequency shift is associated with the
restoring spin force $F_s\propto z$. Roughly speaking, the OSCAR
signal is determined by the ratio $F_{sm}/z_m$, where $F_{sm}$ is
the maximum spin force. When $z_m$ increases, the value $F_{sm}$
approaches its limit $\eta$ (in dimensionless notation). Thus, the
ratio $F_{sm}/z_m$ decreases.

As an example, we estimate the single-spin signal for the
experimental parameters in \cite{1}:
$$
k_c=0.014 {\rm N/m},~\omega_c/2\pi=21.4 {\rm kHz},~ Z_m=28 {\rm
nm},\eqno(12)
$$
$$
~\Bigg|{{\partial B_z}\over{\partial z}}\Bigg|=1.4\times 10^5{\rm
T/m}, ~B_1=0.3 {\rm mT}.
$$
Using Eq. (2) we obtain
$$
\eta=2.9\times 10^{-3},~z_m=8.8\times
10^5,~\varepsilon=390.\eqno(13)
$$
It follows from Eq. (11) that the relative cantilever frequency
shift is $|\Delta\omega_c|/\omega_c=2.3\times 10^{-9}$.

The condition for the adiabatic motion in terms of the
dimensionless parameters
 can be written as
$$
2\eta z_m\ll\varepsilon^2.\eqno(14)
$$
To provide the full reversals of the effective field one also
requires
$$
\varepsilon\ll 2\eta z_m.\eqno(15)
$$
As an example, in the experiment \cite{1} we have: $2\eta
z_m=5.2\times 10^3$, $\varepsilon^2=1.5\times 10^5$. Thus, both
inequalities (14) and (15) are satisfied, i.e. we have full
adiabatic reversals.

Next, we discuss a way to increase the OSCAR signal. It follows
from Eq. (11) that one should sacrifice the requirement for the
full reversals (15) and reduce the amplitude of the cantilever
vibrations $z_m$ and the amplitude of the {\it rf} field
$\varepsilon$, in order to increase the frequency shift. The
minimum possible dimensional cantilever amplitude can be estimated
as
$$
Z_m\sim Z_{rms}=\sqrt{k_BT/k_c}.\eqno(16)
$$
As an example, in the experiment \cite{1}, for $T=3$K, the minimum
value of $z_m$ is $1.72\times 10^3$. Thus, the $z$-component of
the effective field, $2\eta z_m=10$. Taking $\varepsilon=2\eta
z_m$, we still satisfy the adiabatic condition (14) but clearly
violate the condition (15) for full reversals. In this case,
$\Delta\omega_c/\omega_c=7\times 10^{-7}$, and is 300 times
greater than for the experimental parameters (13).

The OSCAR signal should be compared with the frequency thermal
noise $\delta \omega_c/\omega_c$. To estimate this noise we assume
that the noise force changes from zero to its rms value $F_{rms}$
while the $z$-coordinate of the cantilever tip changes from zero
to $Z_m$. Then the effective change of the cantilever spring
constant is $\delta k_c=F_{rms}/Z_m$. Using the expression
$\omega_c^2=k_c/m^*$, where $m^*$ is the effective cantilever
mass, we obtain
$$
{{\delta\omega_c}\over{\omega_c}}={{\delta
k_c}\over{2k_c}}={{F_{rms}}\over{2k_cZ_m}}.\eqno(17)
$$
Finally, using the well-known estimate for $F_{rms}$ (see, for
example, \cite{6})
$$
F_{rms}=2(k_BTk_cB/Q\omega_c)^{1/2},\eqno(18)
$$
where $B$ is the measurement bandwidth, we derive the estimate
for the frequency thermal noise
$$
{{\delta\omega_c}\over{\omega_c}}=(Z_m)^{-1}(k_BTB/k_c\omega_cQ)^{1/2}.\eqno(19)
$$
(This estimate was derived in \cite{7}, using a different
approach.) For the ``natural'' measurement bandwidth,
$B=\omega_c/4Q$, we have from Eq. (19)
$$
{{\delta\omega_c}\over{\omega_c}}=(2QZ_m)^{-1}(k_BT/k_c)^{1/2}.
$$
Taking parameters (12) and $T=3$K, $Q=10^4$, we find
$\delta\omega_c/\omega_c\approx 9.7\times 10^{-8}$, which is
smaller than the estimated OSCAR signal, $\Delta
\omega_c/\omega_c=7\times 10^{-7}$.

Note, that the thermal noise is inversely proportional to the
cantilever amplitude $Z_m$. Thus, the amplification of the OSCAR
signal, associated with a decrease of $Z_m$, does not assume an
increase of the signal-to-noise ratio. However, the decrease of
the field amplitude, $B_1$, which becomes possible for small
values of $Z_m$, would result in reducing the cantilever heating,
which is an important obstacle for a single-spin detection.

So far, we did not discuss the quantum effects in OSCAR. These
effects appear if the initial average spin is not parallel to the
effective field. In the classical case, the angle $\theta$ between
the effective field and the magnetic moment is an integral of
motion. The estimated classical frequency shift
$$
\Delta\omega_c/\omega_c=-{{\eta^2\cos\theta}\over{\sqrt{2\eta^2z_m^2+
\varepsilon^2}}},\eqno(20)
$$
will approach zero as $\theta$ approaches $\pi/2$. In the quantum
case, we expect to obtain one of two possible values in (11)
independent of the initial angle $\theta$. This outcome
corresponds to the standard Stern-Gerlach effect.

\section{Simulations of quantum dynamics}

In our simulations we used the following values of parameters
$$
\eta=0.3,~z_m=13,~\varepsilon=10.\eqno(21)
$$
For the cantilever used in \cite{1} this value of $\eta$
corresponds to a magnetic field gradient of $1.4\times 10^7$ T/m,
and the value of $z_m$ corresponds to the temperature below
$170$$\mu K$.

\begin{figure}
\begin{center}
\includegraphics[width=11cm,height=11cm]{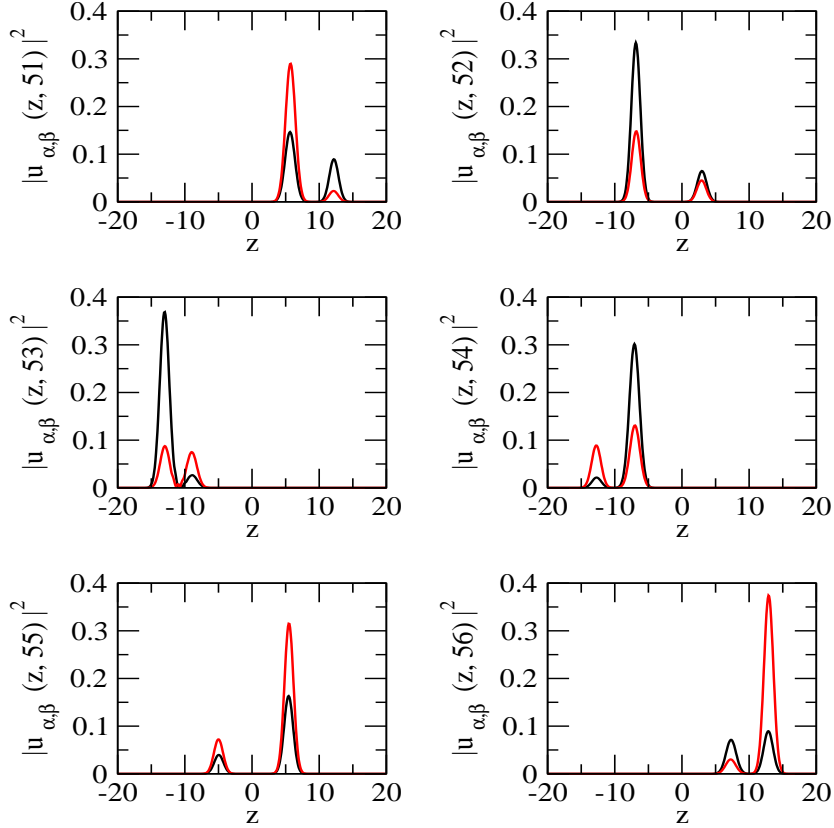}
\vspace{-5mm}
\caption{Functions
$|u_\alpha(z,\tau)|^2$ (black curve) and $|u_\beta(z,\tau)|^2$
(red curve), for six values of $\tau$. Initial conditions:
$\langle z(0)\rangle=z_m=13$, $\langle p(0)\rangle=0$, and the
electron spin points in the positive $z$-direction.} \label{due}
\end{center}
\end{figure}

\begin{figure}
\begin{center}
\includegraphics[height=8cm]{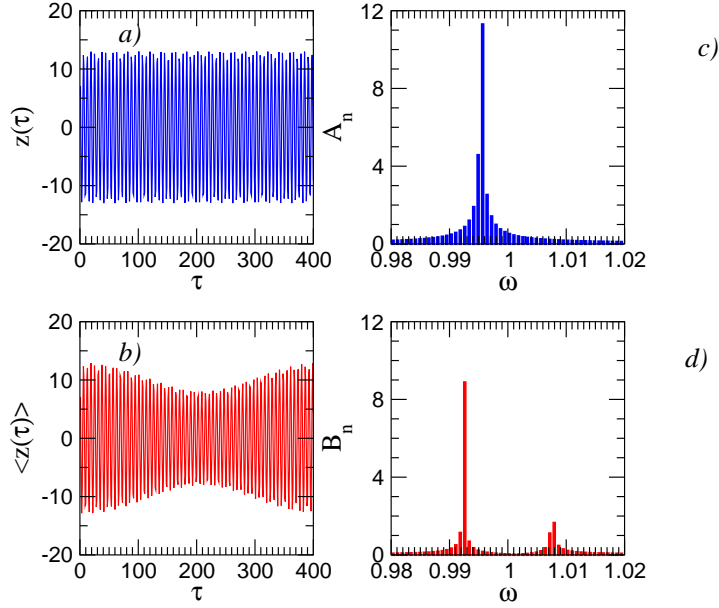}
\vspace{-5mm}
\caption{Classical
and quantum cantilever dynamics in OSCAR MRFM; (a) classical
cantilever coordinate $z(\tau)$, (b) quantum average cantilever
coordinate $\langle z(\tau)\rangle$, (c) the Fourier spectrum for
(a): $z(\tau)=\sum_nA_n\cos(\omega_n\tau+\varphi_n)$, (d) the
Fourier spectrum for (b): $\langle
z(\tau)\rangle=\sum_nB_n\cos(\omega_n\tau+\theta_n)$. All
parameters are the same as in Fig. 2. In a) and b) we show, for
convenience, time sequences shorter than those used in order to
get the Fourier spectrum shown in c) and d). }
\end{center}
\end{figure}

These parameters allow us to reduce the computational time. The
maximum value of the $z$-component of the effective field is,
$2\eta z_m=7.8$. Thus, the condition (14) of adiabatic motion is
satisfied while the condition (15) for full reversals is clearly
violated. The estimated frequency shift is: $|\Delta\omega_c|/\omega_c=7.9\times 10^{-3}$.

The initial conditions which we used correspond to the
quasiclassical state of the cantilever, and the electron spin
pointing in the positive $z$-direction. That means, in Eqs. (4),
$u_\beta (z,0) = 0$ and,
$$
u_{\alpha}(z,0) = \sum_{n=0}^\infty \pi^{1/4}2^{n/2} \alpha_0^n
H_n(z) \exp[-(z^2+|\alpha_0|^2)/2],\eqno(22)
$$
where $H_n(z)$ is a Hermitian polynomial, and
$$
\alpha_0=(\langle z(0)\rangle+i\langle
p(0)\rangle)/\sqrt{2},~\langle z(0) \rangle=13,~\langle
p(0)\rangle=0.\eqno(23)
$$

Our simulations show that the probability distribution for the
cantilever position
$$
P(z,\tau)=|u_\alpha(z,\tau)|^2+|u_\beta(z,\tau)|^2\eqno(24)
$$
eventually splits into two peaks which describe two trajectories
of the cantilever. The ratio of the integrated probabilities for
two peaks is given approximately by $tan^2(\theta/2)$, where
$\theta$ is the initial  angle between the directions of the
average spin and the effective field $\vec
B_{eff}=(\varepsilon,0,-2\eta z(0))$. The approximate position of
the center of the first peak is
$$
z_1=z_m\cos(1-|\Delta\omega_c|/\omega_c)\tau,\eqno(25)
$$
and the approximate position of the second peak is
$$
z_2=z_m\cos(1+|\Delta\omega_c|/\omega_c)\tau,\eqno(26)
$$
where $|\Delta\omega_c|/\omega_c\approx 8.0\times 10^{-3}$. (The
value $|\Delta\omega_c|/\omega_c$ estimated from (11) is almost
the same, $7.9\times 10^{-3}$.) Both functions $u_\alpha(z,\tau)$
and $u_\beta(z,\tau)$ contribute to each peak. (See Fig. 2.) When
two peaks are clearly separated, the wave function $\Psi$ can be
approximately represented as a sum of two functions $\Psi_1$ and
$\Psi_2$, which correspond to two peaks in the probability
distribution. We have found that each function $\Psi_1$ and
$\Psi_2$, with the accuracy to 1\%, can be represented as a
product of the coordinate and spin functions
$$
\Psi_1=R_1(z,\tau)\chi_1(\tau),~\Psi_2=R_2(z,\tau)\chi_2(\tau).\eqno(27)
$$

The first spin function $\chi_1(\tau)$ describes the average spin
which points approximately opposite to the direction of the
effective field $\vec B_{eff,1}=(\varepsilon,0,z_1)$, for the
first cantilever trajectory. The second spin function
$\chi_2(\tau)$ describes the average spin which points
approximately in the direction of the effective field $\vec
B_{eff,2}=(\varepsilon,0,z_2)$, for the second cantilever
trajectory.

Unlike a conventional MRFM dynamics studied in \cite{3,4} the
OSCAR technique implies different effective fields for two
cantilever trajectories.

That is why, in general, the average spins corresponding to two
cantilever trajectories do not point in the opposite directions,
and the wave functions $\chi_1(\tau)$ and $\chi_2(\tau)$ are not
orthogonal to each other. The only exceptions are the instants
$\tau$ for which two cantilever trajectories intersect providing a
unique direction for the effective field.

As we already mentioned, the frequency shift for two cantilever
trajectories $\Delta\omega_c/\omega_c$
 was found to be $\pm 8\times 10^{-3}$. For comparison, we solved the classical equations of motion with
the same parameters and initial conditions as in the quantum case.
The frequency shift for a single cantilever trajectory in the
classical case was found to be $\Delta\omega_c/\omega_c=-5\times
10^{-3}$, which is smaller than the quantum shift and very close
to the value $-5.3\times 10^{-3}$
 derived from the estimate (20). (See Fig. 3.)

\begin{figure}
\begin{center}
\vspace{-5mm}
\caption{Evolution of
the density matrix without the decoherence effects ($Q^{-1}=D=0$),
for four instants of time (as indicated in the figure) .
Parameters: $\varepsilon=10$, $\eta=0.3$. The left column:
$\ln|\sum_s\rho_{s,s}|$, the right column:
$\ln|\sum_s\rho_{s,-s}|$. Initial conditions: $\langle z(0)\rangle
= -8$, $\langle p(0)\rangle = 0$, and the electron spin points in
the positive $z$-direction. Contour lines have been obtained
by intersections of the functions  $\ln|\sum_s\rho_{s,\pm s}|$ 
with  horizontal planes at different heights $h$. Different colours
indicates different regions in the following way :  
$h < -14$ (white) ,$-14< h <-12$ (black) ,
$-12 < h <-10 $ (red), $-10< h <-8$ (green) 
$-8 < h <-6$ (blue) , $-6< h <-4 $ (yellow), 
$h>-4$ (pink) . 
 }
\end{center}
\end{figure}

\section{Decoherence of the Schr\"odinger cat state}

To describe qualitatively the decoherence process, we solved
numerically the master equation (5). The initial density matrix
describes a pure quantum state with the quasiclassical cantilever
and the spin oriented in the positive $z$-direction. At $\tau=0$
we have
$$
\rho_{s,s^\prime}(z,z^\prime,0)=u_\alpha(z,0)u^*_\alpha(z^\prime,0)
\left(\matrix{1&0\cr 0&0\cr}\right),\eqno(28)
$$
where the expression for $u_\alpha(z,0)$ is given in (22), and we
put in (23) $\langle p(0)\rangle=0$, $\langle z(0)\rangle=-8$.
Fig. 4 shows the evolution of the density matrix without effects
of decoherence, for the following values of parameters
$$
\varepsilon=10,~\eta=0.3,~Q^{-1}=D=0.
$$

 The left column demonstrates the behavior of the spin diagonal
density matrix components (contour lines for
$\ln|\sum_s\rho_{s,s}|)$, and the right column demonstrates the
behavior of the spin non-diagonal density matrix components
(contour lines for $\ln|\sum_s\rho_{s,-s}|)$. Initially we have
one peak on the plane $z-z^\prime$. Eventually, this peak splits
into four peaks. Two spatial diagonal peaks, which are
centered on the line $z=z^\prime$, correspond to two cantilever
trajectories. Two spatial non-diagonal peaks describe a
``coherence'' between the two trajectories, which is a
quantitative characteristic of the Schr\"odinger cat state. All
four spin components of the density matrix
$\rho_{s,s^\prime}(z,z^\prime,\tau)$ (with $s=\pm 1/2,s^\prime=\pm
1/2$) contribute to each peak in the $z-z^\prime$ plane.

The density matrix $\rho_{s,s^\prime}(z,z^\prime,\tau)$ can be
represented as a sum of four terms
$$
\rho_{s,s^\prime}(z,z^\prime,\tau)=
\sum_{k=1}^4\rho^{(k)}_{s,s^\prime}(z,z^\prime,\tau),\eqno(29)
$$
where each term describes one peak in the $z-z^\prime$ plane.
Suppose that first two terms in (29) with $k=1,2$ describe the
spatial diagonal peaks, and two other terms with $k=3,4$ describe
the spatial non-diagonal peaks.

We have found that the diagonal terms $\rho^{(1)}$ and
$\rho^{(2)}$ can be approximately decomposed into the tensor
product of the coordinate and spin parts
$$
\rho^{(k)}_{s,s^\prime}(z,z^\prime,\tau)=\hat
R^{(k)}(z,z^\prime,\tau)\hat\chi^{(k)}_{s,s^\prime}(\tau),~k=1,2.\eqno(30)
$$
The spin matrix $\hat\chi^{(1)}_{s,s^\prime}(\tau)$ describes the
average spin which points approximately opposite to the direction
of the effective field $\vec B_{eff,1}$. The spin matrix
$\hat\chi^{(2)}_{s,s^\prime}(\tau)$ describes the average spin
which points approximately in the direction of the effective field
$\vec B_{eff,2}$.

\begin{figure}
\begin{center}
\vspace{-5mm}
\caption{ Evolution of
the density matrix: effects of decoherence and thermal noise;
$D=20$, $Q^{-1}=0.001$, $\varepsilon=10$, $\eta=0.3$. The left
column: $\ln|\sum_s\rho_{s,s}|$, the right column:
$\ln|\sum_s\rho_{s,-s}|$. Initial conditions: $\langle z(0)\rangle
= -8$, $\langle p(0)\rangle = 0$, and the electron spin points in
the positive $z$-direction. 
Contour lines have been obtained
by intersections of the functions  $\ln|\sum_s\rho_{s,\pm s}|$
with  horizontal planes at different heights $h$. Different colours
indicates different regions in the following way :
$h < -14$ (white) ,$-14< h <-12$ (black) ,
$-12 < h <-10 $ (red), $-10< h <-8$ (green)
$-8 < h <-6$ (blue) , $-6< h <-4 $ (yellow),
$h>-4$ (pink).}
\end{center}
\end{figure}

We have found that the same properties of the density matrix
remain for the case $Q^{-1},D\not=0$. Fig. 5 demonstrates the
effects of decoherence and thermal noise for $Q^{-1}=0.001$ and
$D=20$. The spatial non-diagonal peaks in the $z-z^\prime$ plane
quickly decay. This reflects the effect of decoherence: the
statistical mixture of two possible cantilever trajectories
replaces the Schr\"odinger cat state. Next, the spatial diagonal
peaks spread out along the line $z=z^\prime$. This reflects the
classical effect of the thermal diffusion.

\section*{Conclusion}

We presented a quantum theory of the OSCAR MRFM technique. We demonstrated
that the OSCAR signal can be significantly amplified by using partial reversals
 of the effective field instead of the full reversals. If the initial angle
$\theta$ between the directions of the effective field and the
average spin is not small, the quantum frequency shift remains
constant, while the corresponding classical frequency shift is
proportional to $\cos\theta$. This result is a manifestation
 of the Stern-Gerlach effect in the OSCAR MRFM.

Unlike the ``conventional MRFM'' in which the reversals of the
effective field are provided by the frequency modulation of the
{\it rf} field, the OSCAR MRFM exhibits two possible directions of
the effective field. As a result, unlike the conventional MRFM,
and the Stern-Gerlach effect, the two directions of the spin
corresponding to two cantilever trajectories do not remain
antiparallel to each other during the quantum evolution.

\section*{Acknowledgments}

We thank D. Rugar  for helpful
discussions. This work was supported by the Department of Energy
under the contract W-7405-ENG-36 and DOE Office of Basic
 Energy Sciences, by the DARPA Program MOSAIC, and by NSA and ARDA.

{}

\end{document}